\documentclass{article}

\usepackage{microtype}
\usepackage{graphicx}
\usepackage{float}
\usepackage{subcaption}
\usepackage{booktabs}
\usepackage{array}
\usepackage{hyperref}
\usepackage{xcolor}

\usepackage[accepted]{icml2026}

\usepackage{amsmath}
\usepackage{amssymb}
\usepackage{mathtools}
\usepackage{amsthm}
\usepackage[capitalize,noabbrev]{cleveref}

\theoremstyle{plain}

\theoremstyle{remark}

\icmltitlerunning{Affinage: genome-scale mechanistic gene annotation from the published literature}

\begin{document}

\twocolumn[
 \icmltitle{Affinage: Genome-Scale Mechanistic Gene Annotation\\from the Published Literature}

 \begin{icmlauthorlist}
 \icmlauthor{Matteo Di Bernardo}{wi,mit}
 \icmlauthor{Iain M. Cheeseman}{wi,mit}
 \end{icmlauthorlist}

 \icmlaffiliation{wi}{Whitehead Institute for Biomedical Research, Cambridge, MA, USA}
 \icmlaffiliation{mit}{Department of Biology, Massachusetts Institute of Technology, Cambridge, MA, USA}

 \icmlcorrespondingauthor{Matteo Di Bernardo}{matteo.dibernardo@gmail.com}
 \icmlcorrespondingauthor{Iain M. Cheeseman}{icheese@wi.mit.edu}

 \icmlkeywords{LLM, gene annotation, mechanism, UniProt, PubMed,
 biomedical NLP, agentic biology}

 \vskip 0.3in
]

\printAffiliationsAndNotice{}

\begin{abstract}
  Understanding the mechanistic function of a gene is a critical starting point for
  biology. However, for much of the human proteome that knowledge is scattered across
  thousands of primary papers or remains poorly established, while the curated databases
  biologists rely on can lag years behind recent literature. Large language models can now
  read and synthesize that literature on demand, but doing so faithfully for many genes is
  an expensive, non-reproducible retrieval session that does not scale across users. Here,
  we present Affinage, an LLM pipeline that performs this retrieval and mechanistic
  reasoning once per gene --- from the primary literature alone --- and stores the result
  as a reusable, structured annotation. A biologist-designed reading pass extracts only
  direct experimental evidence, and a synthesis pass reasons over those findings alone.
  Applied across the genome, Affinage annotates 19{,}293 human protein-coding genes. This
  analysis provides mechanism for thousands of genes whose UniProt function is empty or a
  stub, beating the curated reference on 99.1\% of head-to-head genes as scored by a
  cross-family LLM judge. Affinage also delineates the $\sim$10\% of the proteome that
  remains mechanistically uncharacterized and will serve as a continuously-updated,
  literature-grounded census of gene function. All records are released openly at
  \url{https://affinage.wi.mit.edu}. More broadly, Affinage serves as an example of 
  how domain experts can encode their expertise into scalable LLM pipelines to improve 
  the publicly available data that guides biological hypotheses and experimentation.
\end{abstract}

\section{Introduction}

\begin{figure*}[t]
\centering
\includegraphics[width=0.92\textwidth]{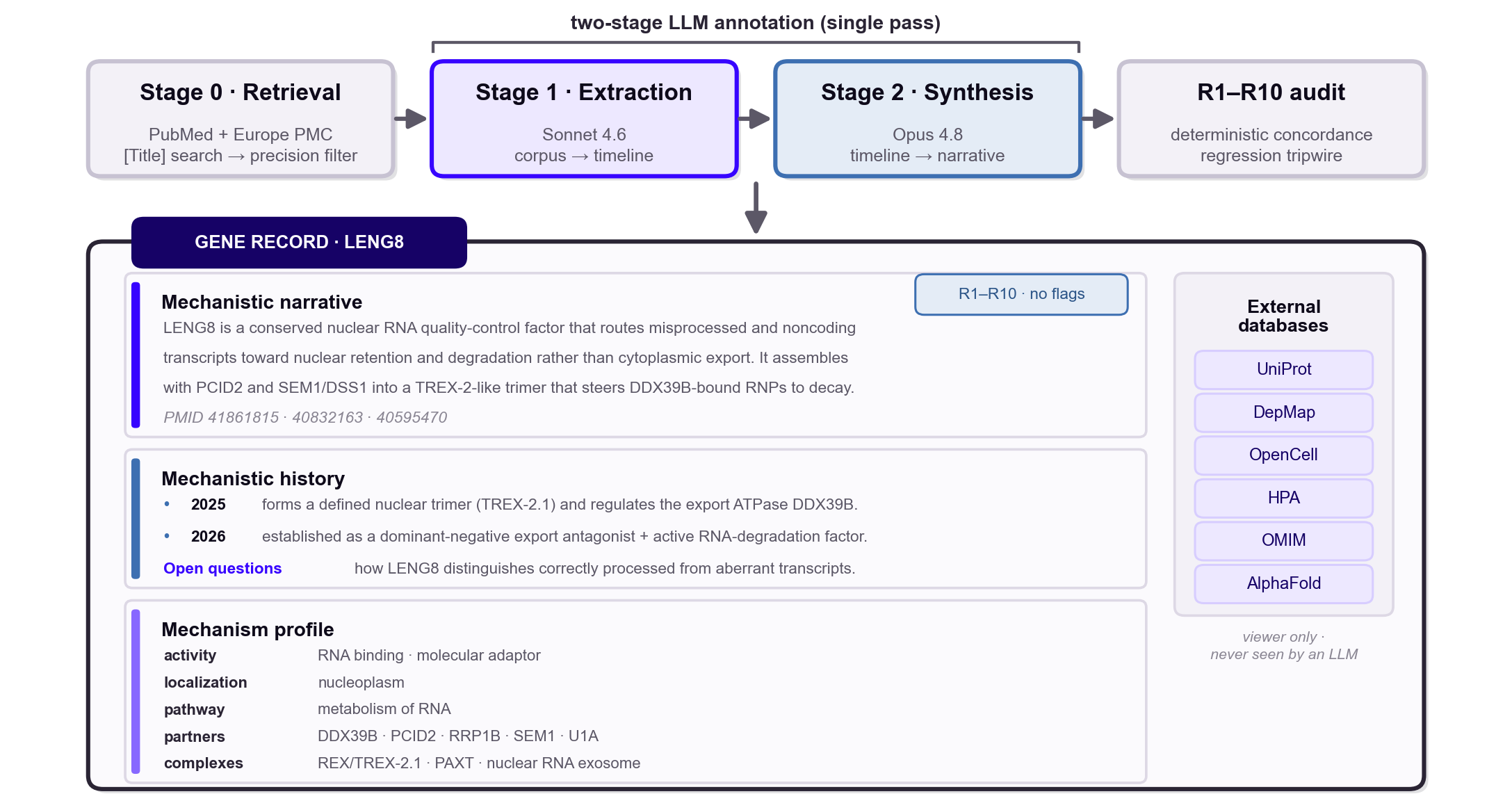}
\caption{\textbf{The Affinage pipeline and the record it produces.}
\emph{Top:} Stage~0 retrieves and precision-filters a per-gene corpus with no
LLM; Stage~1 (Sonnet~4.6) extracts indexed mechanistic findings; Stage~2
(Opus~4.8) synthesizes the narrative from the findings alone; a deterministic
R1--R10 audit screens every record. \emph{Bottom:} each gene yields one
structured record --- a declarative mechanistic narrative, the mechanistic
history (the dated steps and the open questions outstanding at each one), and
the mechanism profile.}
\label{fig:pipeline}
\end{figure*}

A gene's function underlies nearly every experiment or hypothesis a biologist designs, and
yet that knowledge is unevenly recorded: well-studied genes carry rich mechanistic
literature while much of the proteome remains thinly described or not at all. Curated
reference databases often trail primary literature and cannot carry mechanism at the level
reasoning requires. UniProt functional descriptions \citep{uniprot2023} are bound to
manual curation cycles, and Gene Ontology \citep{ashburner2000go,go2023} cannot express
the substrate, structural, or partner-level detail mechanistic reasoning demands. As a
growing ecosystem of computational tools is built on this curated text ---
gene-representation models that embed it as features
\citep{chen2024genept,chen2024genepert} and autonomous bio-AI agents that call it as a
tool \citep{huang2025biomni,mitchener2025kosmos} --- a more continuously updated and
faithful representation of gene function is greatly needed.

Large language models (LLMs) can now read and synthesize the primary literature on demand
\citep{li2025lore}. But naively asking an LLM what a gene does carries three limitations.
First, without carefully designed extraction criteria an LLM overclaims, reporting
phenotypic associations or hypotheses as established mechanism. This erases the
distinction between reasoning over evidence and over noise that a domain expert would
enforce. Second, generation is non-deterministic: the same gene, queried again or by
another user, yields a different answer, so no annotation is stable or reproducible.
Third, grounding each query against the literature or live databases --- increasingly
through per-call tool and MCP access --- is expensive, and that cost compounds across
every gene and every user.

Here, we present Affinage, which performs this retrieval and mechanistic reasoning once
per gene --- under prompts designed by cell biologists to encode what counts as direct
experimental evidence --- and stores each result as a structured, literature-grounded
record any user can query at lookup cost. We release all 19{,}293 human protein-coding
records as a reusable base layer: a biologist can read a gene's mechanism, and a
downstream model or tool ecosystem can ingest it as a richer, more up-to-date substitute
for a UniProt-derived feature, without re-deriving either.

\section{Affinage pipeline}\label{sec:methods}

Affinage is a single-pass, two-stage pipeline (\cref{fig:pipeline}). For each gene, a
deterministic retrieval stage (Stage~0, no LLM) first assembles a per-gene literature
corpus and strips off-target papers. A reading pass (Stage~1) then extracts dated
mechanistic findings, and finally a synthesis pass (Stage~2) reads those findings to
produce the gene's structured record. No curated-database content is ever shown to either
LLM stage, and prefetched reference data is attached to each record for the viewer only.
The deterministic audit layer (\cref{app:audit}) and the LLM-judge validity evaluation
(\cref{sec:eval}) sit outside the generative chain. Full implementation, batch mechanics,
and cost breakdowns are provided in \cref{app:methods}.

\subsection{Stage~0: retrieval and precision filtering}

A biologist exploring a gene's function starts by searching the literature for its name.
We reasoned that, for a holistic representation of a gene's mechanism, a retrieval
strategy would function more faithfully and inexpensively without an LLM in the loop.
Therefore, Stage~0 is a focused retrieval strategy that captures the most relevant papers
for a gene, and avoids pitfalls: that short symbols are ambiguous, that a gene's older
aliases still index real papers, and that a paper merely name-dropping a gene is not
always about it.

Concretely, a title-restricted PubMed search \citep{sayers2022eutils} over the canonical
symbol and its HGNC aliases \citep{seal2023hgnc} assembles the corpus. A precision layer
then drops the off-target papers that a search pulls in, including catalog papers that
name many genes at once, and the literature of a different, unrelated gene whose symbol
happens to collide with an alias. The corpus is then ranked such that the most
gene-specific, well-supported evidence reads first (query construction, short-symbol
disambiguation, and ranking are detailed in \cref{app:methods}). The title-first strategy
is precision-weighted by design. However, a fallback abstract search (triggered when fewer
than 10 papers are found for a given gene) recovers some of these. Even so, not every
paper relevant to a gene is captured. A key paper that never names the gene in its title
can be missed entirely, and because it never enters the corpus the miss is silent, placing
it beyond the reach of the deterministic audit that only checks evidence provided to the
LLM.

\subsection{Stage~1: reading pass}

The reading pass asks of each paper the question a biologist asks when triaging a stack of
abstracts: does this paper actually establish something about how a gene works? Given the
prevalence of noise and tendency to overclaim in biological literature, we designed the
Stage~1 prompt to be strict: the ranked corpus is read by an LLM under a prompt that keeps
a finding only when a direct experiment supports it. It admits substrates from
co-immunoprecipitation or reconstitution, enzymatic activities from in-vitro assay or
active-site mutagenesis, structures with functional validation, pathway positions from
genetic epistasis, post-translational modifications mapped to residues, localization by
imaging or fractionation --- and excludes the rest: phenotypic associations, expression
correlations, and unvalidated computational predictions. It also filters on epistemic
status, so a claim a paper raises and refutes, a hypothesis floated in discussion, or a
mechanism ruled out never enters the record. Each kept finding is stored with its
experimental method, date, journal, and supporting PubMed IDs (PMIDs)
(\cref{app:methods}). When the reading pass returns no findings, a recovery step
distinguishes a genuine absence of mechanistic evidence from a symbol-retrieval failure:
if the gene carries a distinctive UniProt protein name, Stage~0 re-retrieves by that name
--- surfacing literature indexed under the protein rather than the symbol --- and the
corpus is re-read.

\subsection{Stage~2: synthesis pass}

With the relevant findings selected, what remains is the synthesis a biologist would do by
hand: assemble the vetted evidence into a coherent account of what the gene does. In our
experience, decoupling the conclusions of individual papers from a synthesis that reads
only those conclusions is critical to reasoning that faithfully reflects gene function.
Stage~2 reads the evidence layer alone --- never the abstracts --- so it cannot reach for
literature the reading pass did not admit. It returns the gene's record in three layers: a
declarative mechanistic narrative; a per-finding history that carries the open questions
outstanding at the time of each study; and a structured mechanism profile placing the gene
on controlled vocabularies for molecular activity, localization, pathway, named complexes,
and named partners (\cref{app:layers}). The narrative is written under a fixed contract:
declarative and unhedged, with uncertainty carried in the per-finding open questions
rather than the prose, and length scaled to the available evidence so dense genes get more
text and sparse ones are not padded (\cref{fig:s_length}).

\section{Model selection and evaluation}\label{sec:eval}

\begin{figure}[t]
 \centering
 \includegraphics[width=\linewidth]{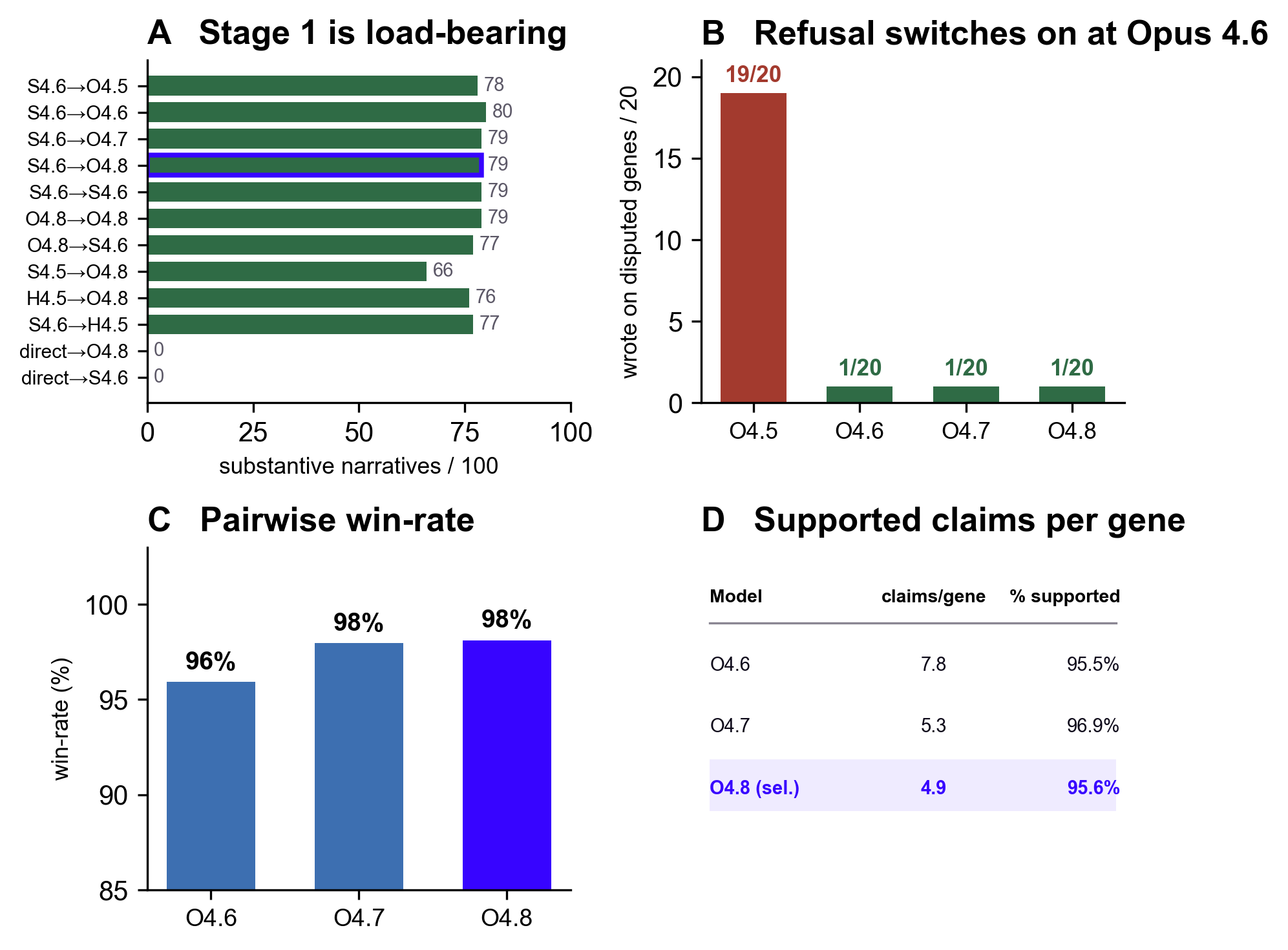}
 \caption{\textbf{Model selection on the 100-gene cohort.} (A)~Stage~1 is
 load-bearing: two-stage configurations produce 77--80 substantive narratives per
 100 genes, while the direct single-stage variants collapse to 0.
 (B)~Refusal switches on at Opus~4.6: on the 20 disputed genes a model should
 decline, Opus~4.5 writes on 19/20 whereas Opus~4.6/4.7/4.8 each write on only
 1/20. (C)~Among the refusal-qualifying models, pairwise win-rate vs.\ UniProt is
 flat, so quality offers no basis to choose among them. (D)~Supported claims per
 gene: the selected Opus~4.8 is the most consolidated (4.9 vs.\ 7.8 for Opus~4.6).
 Faithfulness (0.26\% adjudicated error, 0 contradictions) and the memorization
 probes are reported in the text.}
 \label{fig:eval}
\end{figure}

With this structure in mind, we sought to answer two key questions: (1) How faithfully
Affinage's output traces back to the corpus it reads, and (2) how it improves over what a
curated database like UniProt already offers. We addressed both questions --- and chose
the underlying models behind the pipeline --- on a purpose-built, failure-enriched
100-gene cohort. This cohort includes $\sim$50 of the hardest retrieval and grounding
cases, of which 20 truly have no known mechanism, and $\sim$50 clean controls stratified
by corpus size (\cref{app:cohort}). The refusal and flag rates for this cohort are
deliberately pessimistic and not genome-representative --- the cohort exists to stress the
pipeline and keep configurations commensurable.

After tuning the retrieval Stage~0 over this challenging cohort, we swept the Stage~1 and
Stage~2 generator assignments on this identical cached corpus, isolating the models from
retrieval (\cref{app:ablation}). We found that Stage~1 is load-bearing (\cref{fig:eval}A):
feeding the raw corpus straight to synthesis collapsed output to essentially zero (0/100)
substantive narratives, against 77--80 for the two-stage configurations. The models also
differed in restraint --- whether they declined to write a narrative when the evidence did
not support one, instead of fabricating a plausible mechanism. On the 20 cohort genes
whose retrieved literature establishes no real mechanism (the correct output is no
narrative at all), the older Opus~4.5 wrote a narrative anyway for 19/20, whereas newer
models (Opus~4.6, 4.7, and 4.8) correctly abstained on all but one (\cref{fig:eval}B).

To ground the trust and improvement over UniProt in a non-Anthropic model, we turned to a
cross-family judge (Prometheus-8x7b, a Mistral-family model) that scores pairwise quality
against the UniProt function field (blind, position-swapped) and per-claim faithfulness
against the retrieved corpus (protocols in \cref{app:pairwise,app:faithfulness}). On
quality, the three qualifying models (Opus~4.6/4.7/4.8) tied. Importantly, the win-rate
over UniProt was unanimously high (\cref{fig:eval}C), and per-claim faithfulness was
likewise uniformly high across them (95--97\% supported; \cref{tab:variant-judge}).
Adjudicating every flagged claim on the selected combination against its cited abstracts
left 0.26\% of the 388 cohort claims as genuine errors, with 0 contradictions. Quality
could not break the tie, so we selected the latest combination: Sonnet~4.6
(reading)~$\rightarrow$~Opus~4.8 (synthesis), the most consolidated of the qualifying
combinations (4.9 vs.\ 7.8 supported claims per gene), which beats the UniProt function
field 52/7/1 (wins / ties / losses) on the 60 swap-stable genes (\cref{fig:eval}D).

A final pair of probes checked that the narratives came from the supplied corpus, not from
the model's memorized training data. In the empty-corpus probe, given no corpus at all,
the model must trip the refusal sentinel, which it did successfully on 18/18 genes. In the
wrong-corpus probe, given gene $A$'s identity header but gene $B$'s abstracts, a
memorizing model would recite $A$ and cite papers absent from the supplied corpus; across
18 probes none of the 18 narratives cited an out-of-corpus paper, with 10 grounded in the
supplied ($B$) corpus and 8 refused the mismatch. With selection settled and grounding
confirmed, we applied this configuration across the genome.

\section{Genome-wide results}\label{sec:genome}

\begin{figure}[t]
 \centering
 \includegraphics[width=\linewidth]{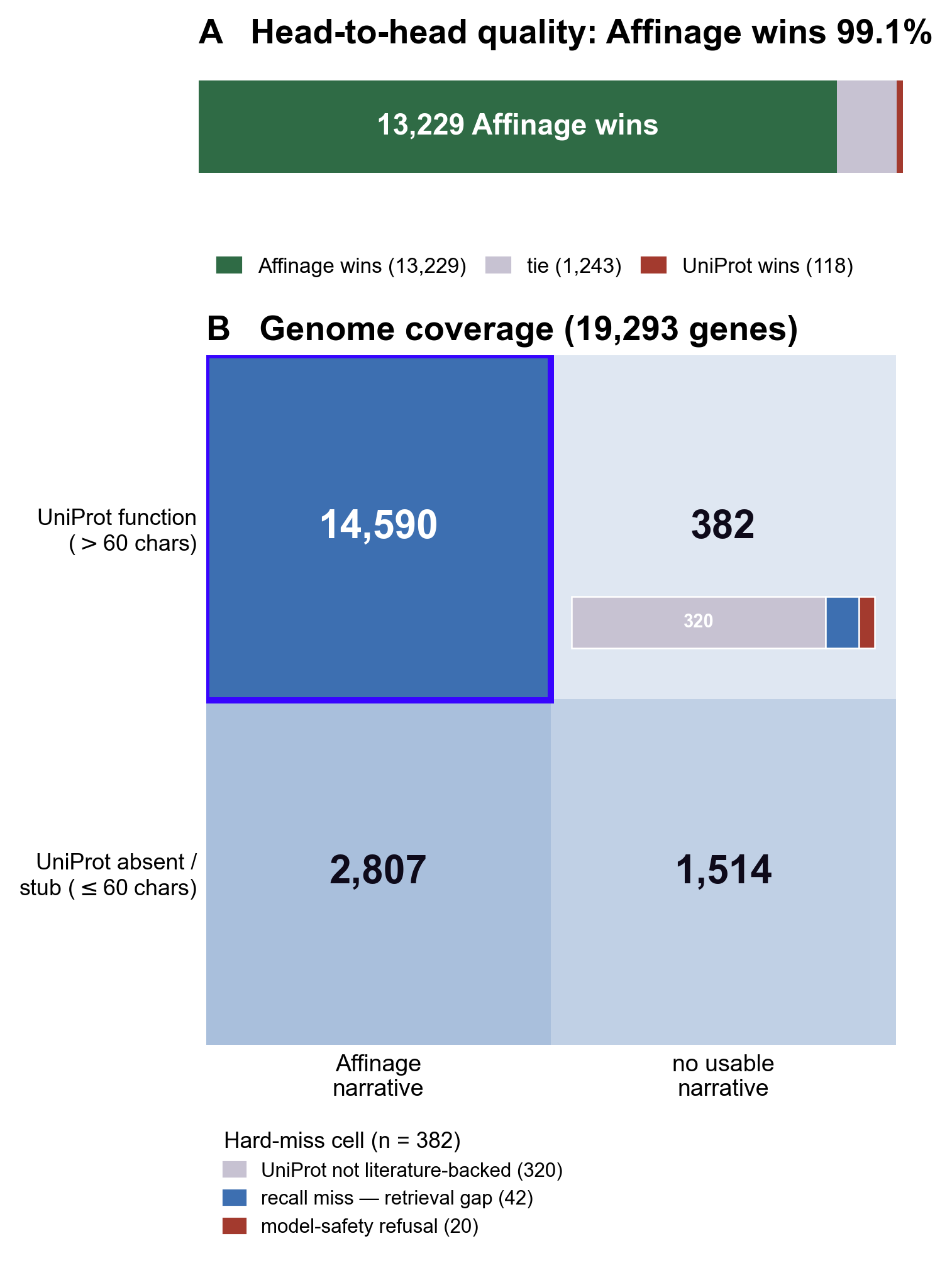}
 \caption{\textbf{Affinage versus UniProt.} (A)~Head-to-head pairwise quality
 (cross-family judge, blind, position-swapped) on the 14{,}590 genes where both
 carry substantive content: Affinage wins 99.1\% of decided pairs. (B)~Coverage
 cross-tab of all 19{,}293 genes by UniProt function ($>60$ chars) vs.\ usable
 Affinage narrative; the violet outline marks the head-to-head set in (A). The
 382 hard misses are further partitioned into 20 model-safety refusals,
 42 recall misses, and 320 correct non-writes where UniProt's own
 function is not literature-backed.}
 \label{fig:coverage}
\end{figure}

\paragraph{Genome-scale resource.}

Run across the genome, the winning Sonnet~4.6~$\rightarrow$~Opus~4.8 configuration yielded
a mechanistic description for nearly every human protein-coding gene --- including biology
too recent or too sparse for curated databases to carry. The released database covers
19{,}293 of 19{,}296 HGNC protein-coding genes (99.98\%): it contains 270{,}143 extracted
mechanism findings (median 11 per gene), open mechanistic questions attached to each
history step, 17{,}399 structured mechanism profiles (90.2\% of annotated genes;
\cref{fig:s_activities}), and 71{,}112 literature-derived partner edges across 14{,}258
genes. The extracted mechanism literature skews recent and grows year on year --- 33\% of
findings cite work from 2020 or later (\cref{fig:s_recency}). The resource is served as a
live REST API and MCP endpoint at \url{https://affinage.wi.mit.edu} (\cref{app:api}).

\paragraph{Genome-scale validation.} To both evaluate and ground the genome-wide run, we
extended the Prometheus pairwise and faithfulness evaluations introduced on the cohort
(\cref{sec:eval}) to every released record. Overall, we found that results in our test
cohort extended to show Affinage consistently matching or beating UniProt. On the 14{,}590
genes where UniProt function text $>60$ characters, the cross-family judge favored
Affinage 13{,}229 / 1{,}243 / 118 (wins / ties / losses) --- a 99.1\% win rate over
decided pairs (\cref{fig:coverage}A). The narratives also stay grounded: run over the
entire resource, the per-claim faithfulness screen scored 97{,}111 claims across 17{,}360
genes at 97.95\% supported and 0.029\% contradicted (28 claims), an unadjudicated 2.05\%
flag rate that upper-bounds the error. Beyond the LLM judge, a large majority of records
also cleared a deterministic structural audit (R1--R10; pure regex and SQL, no LLM): only
206 of 19{,}293 records (1.07\%) tripped at least one flag, most often a cited PMID absent
from the shown corpus, with the per-rule breakdown in \cref{app:audit}.

The genome-wide coverage cross-tab (\cref{fig:coverage}B) splits all 19{,}293 genes by
whether UniProt carries a substantive function and whether Affinage writes a usable
narrative: 14{,}590 carry both (the head-to-head set scored in A), Affinage adds a
narrative for 2{,}807 genes whose UniProt function is empty or a sub-60-character stub
(\cref{fig:s_length}), and 1{,}514 have neither. The 382 genes with a UniProt function but
no Affinage narrative are the apparent misses. We further partitioned these into 20
model-safety refusals, 320 correct non-writes where UniProt's own function is
homology-inferred, unattributed, or curator-only --- not itself literature-backed ---
leaving 42 true recall misses (R5), where UniProt's function is experimentally backed and
on-target literature sat in the retrieved corpus yet nothing was extracted. Across the
genome, Affinage matches or beats the curated reference wherever both exist and extends
mechanism well beyond it.

\section{Limitations: abstract-only vs.\ full-text reading}

The single largest limitation is that Affinage reads abstracts, not full text: mechanistic
detail living only in a paper's methods, results, or supplement is invisible to the
reading pass, and synthesis cannot produce what reading did not extract. The limitation is
bounded, not open-ended --- the faithfulness eval scores any claim overreaching the corpus
as unsupported, so abstract-only reading caps recall but does not silently inflate
precision. However, unlike more complex biological AI tools that seek to dive deep into a
specific biological question, where specific methods, results, and figures may be
critical, we postulated that for a more high-level task, digesting just abstracts would
capture the majority of the mechanism, and that the recall gap would not translate into
quality. To test this, we ran a full-manuscript reading pass on the 100-gene cohort, with
a 1M-token context window to capture as much of the full text as possible. We found that
this task, in itself, was challenging due to known paper paywalls --- we only retrieved
$\sim$46\% of cited papers via PMC, Europe PMC, and Unpaywall. Full text did lift recall
--- $+16.2\%$ more extracted findings (1{,}075 vs.\ 925 on the 76 genes substantively
annotated under both corpora) --- but at $\sim$15$\times$ the Stage-1 token cost, and the
gain did not translate into quality: a cross-family pairwise judge rated the full-text and
abstract-only narratives a statistical tie (20 wins / 33 ties / 24 losses vs.\ the
abstract-only approach). For genome-scale mechanistic annotation, abstract-only reading is
therefore the right default. A retrieval path that targets the specific results sections
carrying novel mechanism is the more promising future direction.

\section{The uncharacterized genome}

Affinage contributes a genome-scale, literature-grounded mechanistic annotation of the
human proteome: one structured record per gene, synthesized only from primary literature,
validated against both the corpus and the UniProt reference, and released openly. It is
most useful on genes that curated databases miss --- those whose mechanism is too recent
or too scattered across the literature to have entered a curated function field. Below, we
provide several example case studies.

LENG8 is one such gene. UniProt lists it as ``Leukocyte receptor cluster member 8'' with
no function entry, and yet all five of its mechanistic papers appeared in 2025--2026.
Affinage resolves LENG8 as a conserved nuclear RNA quality-control factor that, assembled
with PCID2 and SEM1 into the REX complex, acts as a dominant-negative antagonist of
TREX-2, diverting polyadenylated transcripts from nuclear export toward exosomal
degradation. The Affinage record also carries the questions those papers resolved and the
ones still open (e.g.\ the structural basis for recognizing misprocessed transcripts).

Across unrelated biology, Affinage produces equally detailed accounts of gene function. It
resolves KHNYN as a Mn$^{2+}$-dependent endoribonuclease effector of the ZAP antiviral
system that cleaves CpG-enriched viral RNA to restrict HIV-1, SARS-CoV-2, and influenza,
and reconstructs EEPD1's dual role as a structure-specific nuclease channeling stalled
replication forks into homologous recombination and a myristoylated membrane enzyme
activating PKA, tying its loss to cGAS--STING signaling. It identifies CDK19 as a
Mediator-module kinase that also acts kinase-independently in interferon responses,
tracing its \emph{de novo} variants to a syndromic neurodevelopmental disorder with
epileptic features, and assembles ANKRD22's coupling of mitochondrial metabolism to innate
immune signaling through MAVS and NIK. Spanning antiviral defense, replication-fork
repair, transcriptional control, and metabolic--immune coupling, none of these genes
carries a UniProt function field --- each is a new mechanistic description Affinage built
entirely from literature that postdates the curated record.

Importantly, Affinage finds that 1{,}896 human protein-coding genes (9.8\%) still yield no
usable mechanistic narrative. For most genes, this is because the published record does
not yet establish how they work. This residual is itself a result: a continuously-updated,
literature-grounded census of the protein-coding genes that remain mechanistically
uncharacterized, re-derived as new papers appear. Nearly a tenth of the human genome is
still functionally uncharacterized, and that boundary is now trackable gene by gene as the
literature advances.

\section{Conclusions}

Affinage shows that a single-pass, two-stage LLM pipeline --- read the abstract corpus,
then synthesize --- produces genome-scale mechanistic annotation that is both
literature-grounded and validated, not merely fluent. The released resource is one
instance of a broader recipe: a subject-matter expert encodes what counts as evidence into
an LLM pipeline, runs it once at scale, and releases the result openly, turning a one-off
query into a reusable, citation-anchored dataset the field can build on. As LLM-assisted
reasoning becomes routine, this strategy of domain experts using LLMs to produce
high-quality, openly-available data may be among the most powerful ways to raise the
quality of the data the field reasons over.

\section*{Data and code availability}

All gene records, a REST API, and an MCP server are available at
\url{https://affinage.wi.mit.edu}.\footnote{The public endpoint is served from a
warm-pooled deployment with a static fallback of the full record set, so a cold-start
delay does not leave the resource unreachable.} Source code:
\url{https://github.com/cheeseman-lab/affinage}. The reading pass runs Claude Sonnet~4.6
and the synthesis pass runs Claude Opus~4.8; the genome was annotated May~2026 and is
reproducible from the released code, which versions both the extraction and synthesis
prompts. Beyond the source repository, the full extraction, synthesis, and Prometheus
judging prompts are also rendered on the site's About page for direct inspection. The
evaluation environment, including the Prometheus-8x7b judge, is pinned for reproducibility
(vLLM~0.10.2, PyTorch~2.8.0, Transformers~4.55.2), and the validation cohort is seeded.

\section*{Acknowledgements}
We thank members of the Cheeseman laboratory for useful feedback, and Bailey Bova for
discussion and support. This work was supported by a grant from the NIH/NIGMS
(R35GM126930) and the Chan Zuckerberg Initiative (Cell Biology @ Scale) to I.M.C. M.D.\ is
supported in part by an NSF GRFP fellowship (000955563). This research was also supported
by the Whitehead Innovation Initiative (to I.M.C.). LLM inference costs for this project
were funded by the Anthropic AI for Science Program through a compute grant to M.D.\ and
I.M.C.

\bibliography{affinage}
\bibliographystyle{icml2026}

\appendix
\onecolumn

\section{Pipeline and retrieval mechanics}\label{app:methods}

Genome-scale execution chunks the HGNC protein-coding universe \citep{seal2023hgnc} into
batches of $\le 2{,}000$ genes, sized by the batch API's payload limit ($\sim$256~MB)
against the observed mean of $\sim$80~KB of formatted prompt per gene; larger chunks are
rejected at the edge. Each chunk runs sequentially through retrieval, the reading pass,
and the synthesis pass, then upserts into the database. Chunks are independent: a failed
chunk does not block the next, and because the upserts are idempotent an interrupted run
can simply be relaunched.

Retrieval is the wall-time long pole. Each gene's corpus and reference metadata are
disk-cached, and a gene whose cache already exists is skipped, so partial reruns reuse
prior PubMed \citep{sayers2022eutils} and iCite \citep{hutchins2016rcr} work rather than
re-querying the APIs. Fresh retrieval runs at $\sim$50~minutes per 2{,}000-gene chunk and
a warm cache reduces this to $\sim$5~minutes, making interrupted-run restarts effectively
free. All NCBI requests share a single rate limiter that holds the published request
budget across the parallel retrieval and prefetch workers, and per-gene failures
(transient NCBI errors, schema-invalid synthesis output) are logged and skipped so the run
continues.

The title query searches the canonical symbol plus up to five HGNC aliases (covering both
previous and current gene symbols) and is filtered by biological terms to suppress
non-biomedical hits; short symbols ($\le$4 characters) receive additional disambiguated
queries (``\{gene\} protein'' and the UniProt full name), since short names collide with
common abbreviations, and if the title search returns fewer than ten papers it broadens to
title-or-abstract, recovering recall for genes too sparsely studied to be named in a
title. bioRxiv/medRxiv preprints are added via Europe PMC \citep{europepmc2015}. The
precision layer's catalog denylist is seeded with 262 PMIDs that each appear in more than
50 genes' corpora (8.2\% of corpus slots) and is recomputed per run to track new catalog
papers; for a gene whose alias is itself another gene's canonical symbol, the query is
restricted to the canonical plus previous symbols, stopping the collision at the fetch
step. Survivors are ranked by inverse gene fan-out, then by whether the gene is the
title's primary subject, with peer-review status and iCite citation count as tiebreaks;
HGNC serves here only as a naming authority, never as a curation set.

Each finding the reading pass keeps is recorded as a structured evidence entry with a
two-axis confidence score --- method tier $\times$ evidence preponderance --- so the
synthesis pass can weight stronger evidence over weaker.

\paragraph{Index-based citation.} The synthesis pass cites each claim by the integer
index of the finding it rests on, never by writing out a reference identifier; a
deterministic post-processing step then resolves each index to the PMIDs recorded for that
finding. Because the model never emits a citation token along this path, a confabulated or
truncated reference cannot arise from index resolution --- removing the dominant failure
mode of literature-grounded generation for every citation routed through the index. The
residual, a raw PMID emitted outside the index path, is what the audit layer monitors
(\cref{app:audit}).

\paragraph{Incremental maintenance.} The pipeline supports an incremental-rerun mode for
keeping the resource current as new literature appears. On a rerun, retrieval re-fetches
each gene's corpus rather than serving it from cache, then compares the new reference set
against the set stored on the prior record: genes with no new papers are copied forward at
zero model cost, and only genes with at least one new paper re-enter the reading and
synthesis passes. Idempotent upserts let an interrupted maintenance pass be relaunched
without duplicating work.

\paragraph{Cost.} Costs are reported at the batch rate (50\% of on-demand;
\citealt{anthropic2025claude}). Synthesis dominates spend: the Opus synthesis pass costs
$\sim$2.4$\times$ the per-gene reading pass, with the per-gene total amortizing to a mean
of \$0.13 (median \$0.11) and a total genome-scale spend of \$2{,}505. The deterministic
audit layer (\cref{app:audit}) calls no API and runs in $\sim$30~seconds over the
read-only database.

\section{Deterministic audit layer}\label{app:audit}

The audit layer is a deterministic regression tripwire, not a measure of semantic
correctness: it catches structural failures that should not survive the single-pass
pipeline. Every rule is pure regex and SQL with no LLM in the evidentiary chain, reading
only the narrative text, the HGNC alias table, the extracted finding timeline, and the
length of the prefetched UniProt function field --- never any curated-database content,
preserving the database-exclusion invariant. The ten rules fall in three tiers;
\cref{tab:audit} gives what each catches and how often it fires across the released
resource, and the full per-rule specifications are published alongside the resource
(\url{https://affinage.wi.mit.edu}).

The IDENTITY tier (R1--R4) catches a narrative about the wrong gene, or the wrong product
of the right gene: an alias that is itself another gene's canonical symbol dominating the
text, an opening led by a non-human organism homonym, a first noun phrase naming a
different symbol, or an opening dominated by a non-coding product of the locus.

The GROUNDING tier (R5--R8) catches a narrative that under-extracts or misuses evidence: a
refusal placeholder despite substantive UniProt function text, citations that barely
overlap the gene's independently indexed literature (a corroborating signal only), a cited
reference absent from the shown corpus (a truncation or fabrication), or too large a
fraction of uncited body claims.

The BEHAVIOR tier (R9--R10) catches model-behavior anomalies: a failed contamination
probe, or no usable narrative at all --- a model-safety refusal, a parse failure, or an
over-refusal. A placeholder with zero findings is the correct outcome for an empty
timeline and is never flagged.

R7 is the most frequently tripped rule (66 records, 0.34\%): a PMID cited in the narrative
but absent from the shown corpus. In these cases the underlying claim is grounded in a
real corpus paper, but the rendered identifier is mis-attributed or truncated --- a
citation-grounding failure, not a wrong-gene error. The next is R5 (42), the deterministic
recall miss --- a gene with experimentally-backed UniProt function and on-target corpus
evidence that yielded no narrative; R8 flags a further 28 uncited-synthesis cases. The
IDENTITY tier (R1--R4) flags 46 records and BEHAVIOR (R9--R10) 27, the latter (R10,
0.14\%) dominated by model-safety refusals enriched for host--pathogen and viral-entry
genes.

\begin{table}[h]
 \caption{\textbf{Deterministic structural audit (R1--R10).} What each rule
 catches and its flag count across the released genome-scale resource;
 reproducible from the database alone, no LLM. Counts are per rule, so a gene
 tripping several rules is counted under each; R6 only corroborates an IDENTITY
 call and never escalates alone.}
 \label{tab:audit}
 \vskip 0.05in
 \centering
 \footnotesize
 \setlength{\tabcolsep}{4pt}
 \renewcommand{\arraystretch}{1.20}
 \begin{tabular}{@{}l l l r@{}}
 \toprule
 \textbf{Tier} & \textbf{Rule} & \textbf{Fires when} & \textbf{Genes}\\
 \midrule
 IDENTITY  & R1 & alias dominates the narrative \emph{and} the contamination reaches it & 26\\
           & R2 & opening led by a non-human organism qualifier & 0\\
           & R3 & opener names or equates the gene to a different HGNC symbol & 11\\
           & R4 & opening dominated by a non-coding product of a protein-coding locus & 9\\
 \midrule
 GROUNDING & R5 & no narrative despite experimental UniProt + on-target corpus & 42\\
           & R6 & citations barely overlap the gene's indexed literature & 8\\
           & R7 & cited PMID absent from corpus (truncation / fabrication) & 66\\
           & R8 & uncited body-claim fraction $> 0.30$ & 28\\
 \midrule
 BEHAVIOR  & R9  & contamination probe failed (empty- or wrong-corpus) & 0\\
           & R10 & no usable narrative (refusal / parse-failure / over-refusal) & 27\\
 \midrule
 \multicolumn{3}{l}{\textbf{Records flagged (any rule)}} & 206 (1.07\%)\\
 \bottomrule
 \end{tabular}
\end{table}

\section{Faithfulness and pairwise evaluation}\label{app:eval}

Both the faithfulness and pairwise judgements use a deliberately cross-family LLM judge
(Prometheus-8x7b, Mistral family) so there is no self-preference bias toward the Anthropic
generators; the judge never sees any curated database, matching the pipeline's own
database-exclusion invariant. The full judging rubrics, like the extraction and synthesis
prompts, are published alongside the resource (\url{https://affinage.wi.mit.edu}).

\subsection{Faithfulness rubric}\label{app:faithfulness}

In the absolute (faithfulness) mode the judge reads each decomposed narrative claim with
the gene's retrieved corpus and scores it on a three-level rubric (supported / unsupported
/ contradicted); open-question claims are routed out deterministically before scoring. An
automated judge over-flags this task --- it effectively scores a multi-cited claim against
only the first of its cited abstracts and treats a paper indexed under a gene synonym as a
``different gene'' --- so we use it as a high-recall screen and manually adjudicate every
flagged claim against the full set of its cited abstracts. On the cohort almost all flags
resolve to supported once that full set is read (e.g.\ IMP4/Mpp10, ZKSCAN1, STAG1, GTF3C2,
TTC7B); the rare genuine error is a paralog or gene-family mis-attribution introduced at
retrieval (e.g.\ SLC16A4 carrying SLC16A3/MCT4 glycolytic biology, its corpus legitimately
co-mentioning the SLC16 family), and contradictions are absent. The screen's run-to-run
movement is itself within per-claim judge noise, whereas the adjudicated error rate is
stable across resamples --- which is why the adjudicated rate, not the raw screen, is the
headline. At genome scale the same screen runs unadjudicated; the contradiction rate,
which needs no adjudication, is the load-bearing signal there.

\subsection{Pairwise protocol}\label{app:pairwise}

The relative (pairwise) mode presents the judge with two descriptions of the same gene and
asks which is better supported by the retrieved corpus --- here the Affinage narrative
against the UniProt function field, on genes carrying both. Each pair is source-blind (the
judge is not told which side is Affinage) and scored under both presentation orders
(position swap) to control for order bias; a verdict counts only when it is stable across
the swap. The win/tie/loss outcome is reported in the body and \cref{fig:eval}B.

\subsection{Validation cohort}\label{app:cohort}

Model selection and validation use a purpose-built, failure-enriched 100-gene cohort
(seeded for reproducibility). Failure-class genes are drawn by audit-subtype quota (alias
collision, alternate product, paralog, recall miss, uncited synthesis) plus over-refusal
probes; the clean controls are stratified by corpus size (sparse $<5$, moderate 5--20,
rich $>20$ papers). The identical set runs through every configuration and comparison.
Re-classified against the current audit rules, the realized composition is 57
failure-class and 43 control genes (a few build-time controls trip a current rule).

\subsection{Model and architecture ablation}\label{app:ablation}

The full per-variant sweep summarized in \cref{fig:eval} crosses the reading- and
synthesis-stage generator assignments on the identical cohort and cached corpus, holding
the database-exclusion invariant fixed; the deterministic audit-layer counts (substantive
narratives, refusals, structural flags) are the discriminating selection axis.
\cref{tab:variant-judge} reports the cross-family judge on the same configurations. The
configurations cluster tightly --- supported rates and pairwise wins over UniProt span
only a few points, within the per-claim judge noise --- so the judge corroborates rather
than drives the choice. Selection rests on the deterministic axes (refusal discipline and
contract adherence), where the winning combination is clearly strongest and also carries
the joint-lowest adjudicated error rate.

\begin{table}[h]
 \centering
 \small
 \caption{\textbf{Cross-family judge across configurations} (100-gene cohort).
   The supported-rate column is the per-claim screen before manual adjudication
   (see \cref{app:faithfulness}); the vs.\ UP column is pairwise win/tie/loss
   against UniProt over swap-stable genes. Differences are within judge noise;
   the configuration is selected on the deterministic axes.}
 \label{tab:variant-judge}
 \begin{tabular}{lcc}
 \toprule
 \textbf{Stage~1 $\to$ Stage~2} & \textbf{Supp.\,\%} & \textbf{vs.\ UP}\\
 \midrule
 S4.6 $\to$ O4.8 (winning) & 95.6 & 52/7/1 \\
 S4.6 $\to$ O4.5 & 94.8 & 51/11/6 \\
 S4.6 $\to$ O4.6 & 95.5 & 47/12/2 \\
 S4.6 $\to$ O4.7 & 96.9 & 48/11/1 \\
 S4.6 $\to$ S4.6 & 96.8 & 52/7/1 \\
 S4.6 $\to$ H4.5 & 96.0 & 53/5/1 \\
 S4.5 $\to$ O4.8 & 97.7 & 44/6/1 \\
 H4.5 $\to$ O4.8 & 96.7 & 52/6/1 \\
 O4.8 $\to$ O4.8 & 98.1 & 54/5/2 \\
 O4.8 $\to$ S4.6 & 97.8 & 48/11/1 \\
 \bottomrule
 \end{tabular}
\end{table}

\section{The resource}\label{app:resource}

\subsection{Output layers}\label{app:layers}

Each gene record exposes three synthesis layers, all produced by the synthesis pass from
the extracted findings alone. The mechanistic narrative is the declarative synthesis
described in the body --- the closest analog to a UniProt function field, but with
per-finding citation provenance and recent biology integrated. Its length is calibrated to
the empirical distribution of UniProt function-field lengths rather than a fixed cap, so
output scales with the depth of available evidence. The mechanistic history is a
per-finding record of what each finding established, together with an explicit gaps field
carrying the open questions outstanding at the time of each finding --- the sole home for
uncertainty, which is banned from the narrative. The mechanism profile places each gene on
controlled vocabularies across the molecular-activity, localization, and pathway axes.

\subsection{API and MCP access}\label{app:api}

The resource is served as a live public REST API and an MCP server at
\url{https://affinage.wi.mit.edu}, exposing per-gene records (all three synthesis layers
plus the attached reference data) for programmatic and agent access. The deterministic
audit flags are surfaced per gene through a dedicated endpoint and MCP tool, so a consumer
can read the structural-QC status alongside the annotation. Symbol lookups are
alias-aware: a query against a previous or non-canonical symbol resolves to the canonical
record via HGNC (for example, ZNRD1 redirects to POLR1H), so callers reach the right gene
regardless of which historical symbol they hold.

\clearpage
\section{Supplementary figures}\label{app:suppfigs}
\setcounter{figure}{0} \renewcommand{\thefigure}{S\arabic{figure}}

\begin{figure}[H]
\centering
\includegraphics[width=0.92\textwidth]{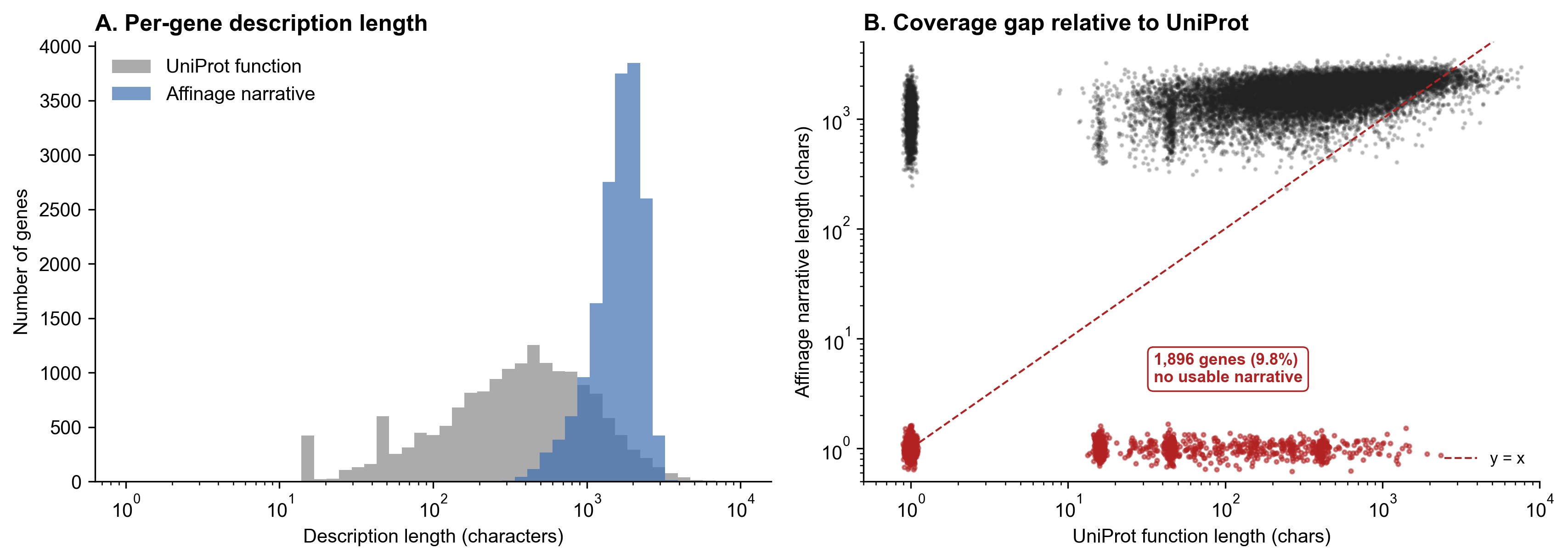}
\caption{\textbf{Per-gene description length, Affinage versus UniProt.}
(A)~Length distributions over genes with usable content. (B)~Per-gene scatter;
the 1{,}896 genes with no usable narrative (9.8\%) are floored to $y=1$ (red).}
\label{fig:s_length}
\end{figure}

\begin{figure}[H]
\centering
\includegraphics[width=0.92\textwidth]{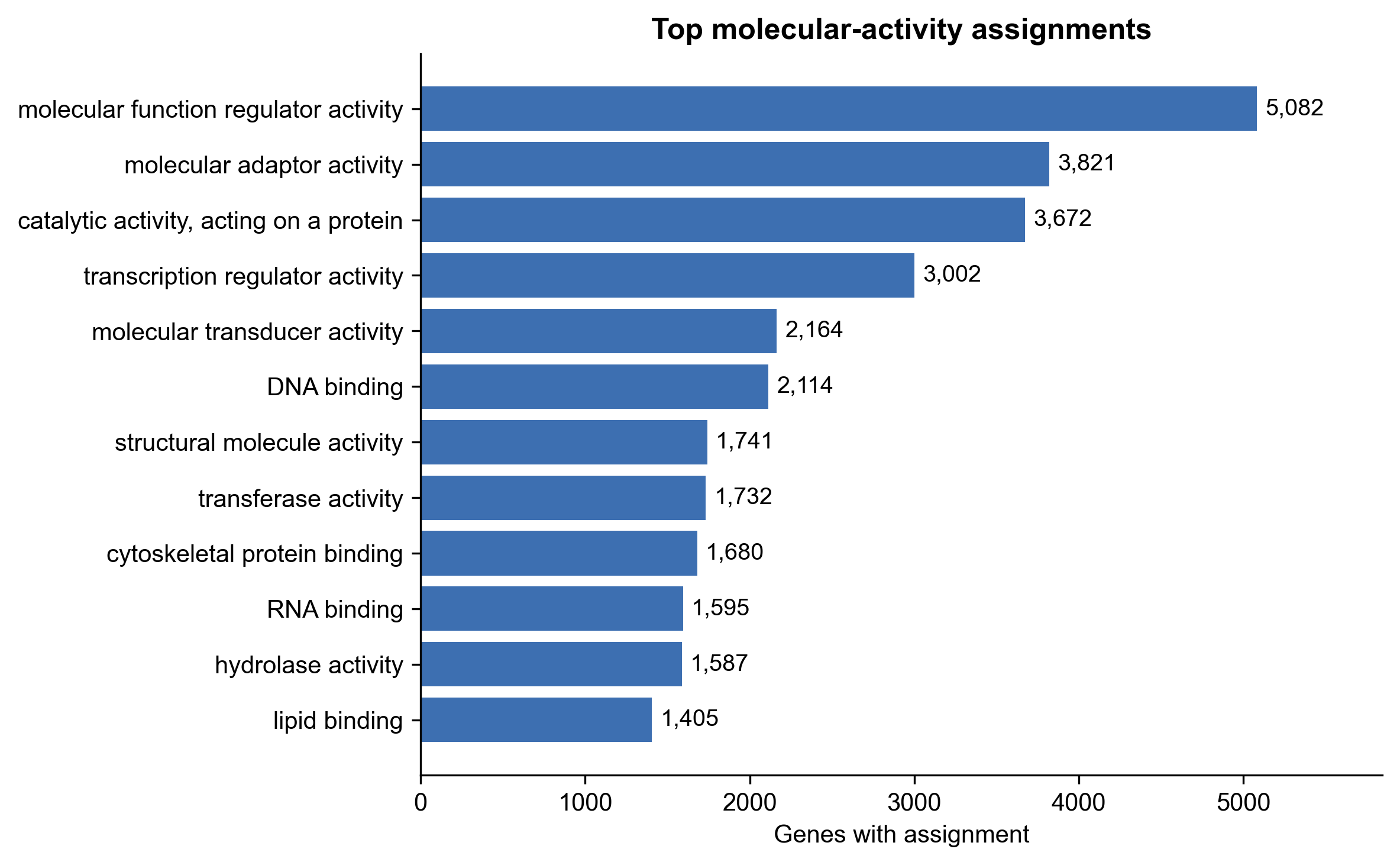}
\caption{\textbf{Most frequent molecular-activity terms.} Top
controlled-vocabulary molecular-activity assignments across the resource;
``molecular function regulator activity'' leads at 5{,}082 genes.}
\label{fig:s_activities}
\end{figure}

\begin{figure}[H]
\centering
\includegraphics[width=0.92\textwidth]{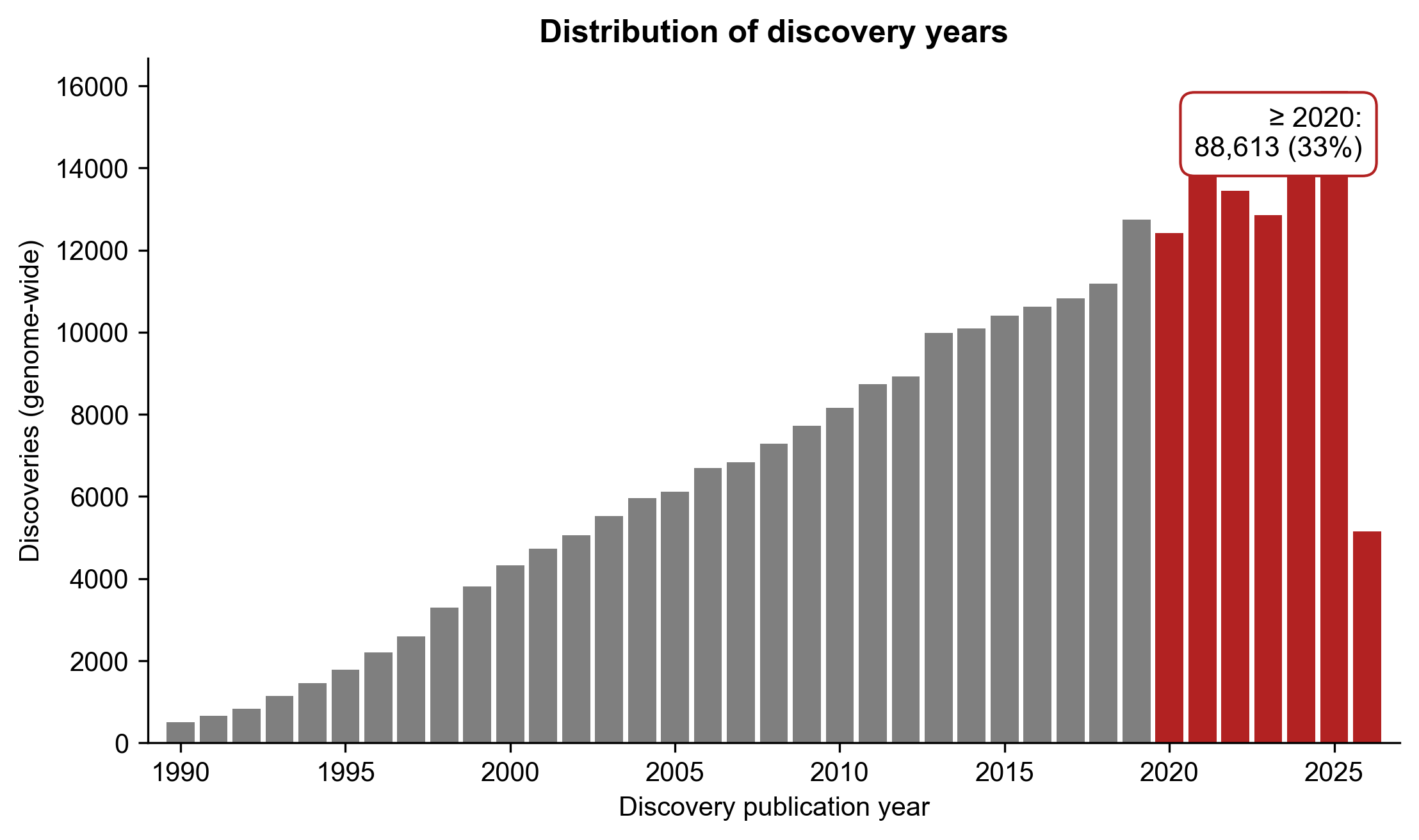}
\caption{\textbf{Recency of the extracted mechanism literature.} Publication-year
distribution across the 270{,}143 extracted findings; 32.8\% (88{,}613) cite
literature from 2020 or later.}
\label{fig:s_recency}
\end{figure}

\end{document}